\documentclass[prl,twocolumn,showpacs,preprintnumbers,superscriptaddress]{revtex4}
\usepackage{amsfonts}

\usepackage{eurosym}
\usepackage{color,graphicx}
\usepackage[dvips]{epsfig}

\begin{document}

\title{A micrometer-scale integrated silicon source of time-energy entangled photons}
\author{Davide Grassani}
\author{Stefano Azzini}
\author{Marco Liscidini}
\author{Matteo Galli}
\affiliation{Dipartimento di Fisica, Universit\`{a} degli Studi di Pavia, via Bassi 6,
27100 Pavia, Italy}
\author{Michael J. Strain}
\affiliation{School of Engineering, University of Glasgow, Glasgow G12 8LT, UK}
\affiliation{Institute of Photonics, University of Strathclyde, Glasgow G4 0NW, UK}
\author{Marc Sorel}
\affiliation{School of Engineering, University of Glasgow, Glasgow G12 8LT, UK}
\author{J. E. Sipe}
\affiliation{Department of Physics and Institute for Optical Sciences, University of
Toronto, 60 St. George Street, Ontario, Canada.}
\author{Daniele Bajoni}
\email{[]daniele.bajoni@unipv.it}
\affiliation{Dipartimento di Ingegneria Industriale e dell'Informazione, Universit\`{a}
degli Studi di Pavia, via Ferrata 1, Pavia, Italy}
\date{\today}



\begin{abstract}

Entanglement is a fundamental resource in quantum information processing. Several studies have explored the integration of sources of entangled states on a silicon chip but the sources demonstrated so far require millimeter lengths and pump powers of the order of hundreds of mWs to produce an appreciable photon flux, hindering their scalability and dense integration. 

Microring resonators have been shown to be efficient sources of photon pairs, but entangled state emission has never been demonstrated.
Here we report the first demonstration of a microring resonator capable of emitting time-energy entangled photons. We use a Franson experiment to show a violation of Bell's inequality by as much as 11 standard deviations. The source is integrated on a silicon chip, operates at sub-mW pump power, emits in the telecom band with a pair generation rate exceeding 10$^7$ Hz per $nm$, and outputs into a photonic waveguide. These are all essential features of an entangled states emitter for a quantum photonic networks.

\end{abstract}

\maketitle

\section{Introduction}

Photonics is increasingly seen as an attractive platform for quantum
information processing \cite{NielsenandChuang,Ekert,cryptoent,Ent_Role}. In quantum cryptography \cite{Ekert,BB84} photons have several advantages as vectors of information, due to their long coherence times at room temperature and the possibility of being transmitted over the existing optical fiber infrastructure. 
The potential scalability and integrability of photonics
also suggests its application in quantum simulation and computing \cite{Osellame,Metcalf,Shadbolt,Silverstone}.
The most common strategy for producing entangled photon pairs at room temperature is the use of the parametric fluorescence that can occur in a non-linear crystal \cite{EP12,Fujii}. While having high generation rates, these sources are very difficult to integrate. 
An ideal integrated source of entangled photons 
should be CMOS compatible for cost-effective and reliable 
production, easily interfaced with fiber networks for long range
transmission in the telecom band, and take up little ``real estate'' on the chip. In these regards, the main results have been obtained by exploiting third order nonlinearities in silicon, and they are focused on the generation of qubits based on polarization entangled photon pairs \cite{Take_Pol,Olislager} or entangled time-bins \cite{Take_0,Take,Take_scirep} in long waveguiding structures. However, these devices require lengths ranging from fractions of a millimeter to centimeters to produce an
appreciable photon pair flux, hindering their scalability.
Another kind of quantum correlation of photon pairs is \emph{time-energy} entanglement. This is arguably the most suitable format for the entanglement, as it can be easily manipulated in integrated optical circuits \cite{Shadbolt}, and it can be preserved over long distances in the fiber optical networks \cite{Tittel,Marcikic} needed for communication between devices. Very recently, it has been shown that the use of time-energy entangled photon pairs in quantum key distribution can enable higher key generation rate compared to entangled photon pairs in lower-dimensional Hilbert spaces \cite{EnglundPRA2013}.

In this work we demonstrate that silicon ring resonators in a silicon-on-insulator platform 
are an efficient source of time-energy entangled photon pairs. 
The large field enhancements that
can be obtained in  resonant structures \cite{Davanco,PhM} and ring resonators in particular \cite{Lipson_nat,Clemmen} combined with the large effective nonlinearities achievable in
silicon ridge waveguides, of which
they are made, allows the reduction of the emitter's footprint by orders of magnitude and the drastic improvement of the wavelength conversion efficiency, together with the spectral properties of the emitted pairs, with respect to silicon waveguide sources.

\section{Sample structure and transmission spectra}

The sample geometry is illustrated in Fig. 1a: the device is a ring
resonator with a radius of 10 $\mu$m, evanescently coupled to a straight
silicon waveguide on one side of the ring; both the ring and the waveguide
have transverse dimensions of 500 nm (width) and 220 nm (height) and
are etched on a silicon-on-insulator wafer. The gap between the ring and the
waveguide is 150 nm. The coupling of light onto and off of the chip is
implemented by mode field converters, and the emission is extracted through
a tapered optical fiber (Supplementary Fig. 1 and Supplementary Note 1). A tunable continuous wave laser is used for
characterizing the sample, and as a pump for the nonlinear optical
experiments (see Methods). While such ring resonators would act as all-pass devices in the absence of scattering losses, they are somewhat akin to integrated Fabry-Perot cavities in that the modes of the electromagnetic field are identified by a comb of resonances. The transmission spectrum from our sample is shown in Fig. 1b, where the dips occur due to scattering losses at the resonances.  The free spectral range is about 9 nm, and the resonance quality factors (Qs) are, on average, around 15000. The minimum transmission
is about 3-5\% on resonance, meaning that the ring almost satisfies the critical coupling condition, which maximizes the coupling
between the ring and the bus waveguide.

\begin{figure}[h!]
\includegraphics[width= \columnwidth]{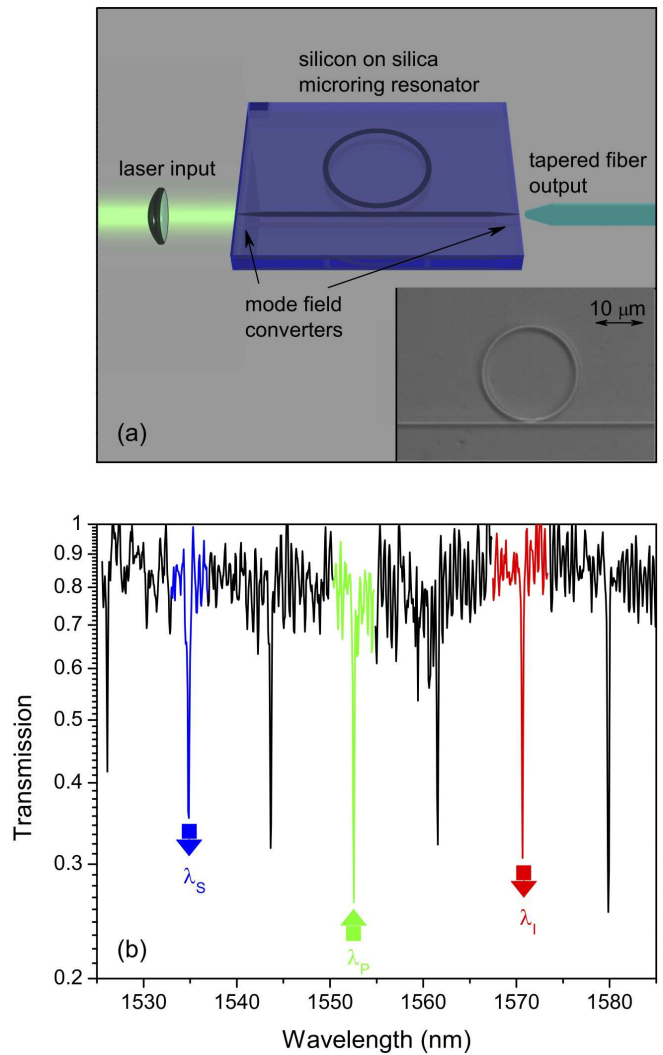}
\caption{\textbf{Sample structure and characterization.} (a) Sketch of the sample together with the input/output light
coupling mechanism. The R=10 $\protect\mu$m ring resonator is evanescently
coupled to a silicon nano-wire waveguide via a deep-etched 150 nm gap point
coupler. An optical microscope image of the ring is shown in the inset. The
waveguide ends at both sides with spot-size converters: 300 $\protect\mu$m
long silicon inverse tapers ending in a 20 nm tip width covered by 1.5x2.0 $%
\protect\mu$m$^2$ polymer waveguides. Light injection from a collimated pump
laser is achieved by the use of an aspheric lens with numerical aperture
NA=0.5. The output from the sample is collected with a PM lensed-fiber with
a working distance of 3 $\protect\mu$m. (b) Transmssion spectrum of the
resonator. The pump resonance is highlighted in green, the signal and idler
resonances employed in the experiment are indicated in blue and red,
respectively. }
\label{Fig1}
\end{figure}

\section{Nonlinear spectroscopy and coincidence measurements}

The nonlinear process responsible for the generation of photon pairs is spontaneous four wave mixing \cite{Clemmen,Sharping,Kraus,OE,Engin}: two pump photons at frequency $\omega _{p}$ are converted into signal and idler photons at frequencies $\omega _{s}$ and $\omega _{i}$ (as sketched in the
inset of Fig. 1b). When using resonant structures, energy conservation implies three equally spaced resonances in energy ($
\hbar \omega _{s}+\hbar \omega _{i}=2\hbar \omega _{p}$). Another advantage in using ring resonators is that the process is greatly amplified by the resonance, and it has been shown \cite{OL} that the generation rate goes as $R\propto \frac{Q^3}{R^2}P^2$ where $Q$ is the quality factor of the resonances, $R$ the ring's radius and $P$ the pump power. In our ring
resonators, waveguide dispersion limits the bandwidth over which pairs can be generated for a fixed pump wavelength to a spectral
range of about 80 nm, resulting in plentiful choice of possible signal and idler pairs: we have also verified that generation rate is almost the same for the fourth neighbouring resonances from the pump \cite{OL}. We note that in fact a single pump pulse will generate a number of entangled signal and idler pairs in parallel, with each entangled pair easily extracted because of their frequencies. But in this work we study only one pair, as highlighted by colors in Fig 1b: we use a resonance around 1550 nm (at the center of the telecommunication c-band) for the pump and its second nearest neighbor resonances for signal and idler; this spectral distance is chosen to optimize filtering of the laser
background noise at the signal and idler frequencies.

\begin{figure*}[t!]
\includegraphics[width= \textwidth]{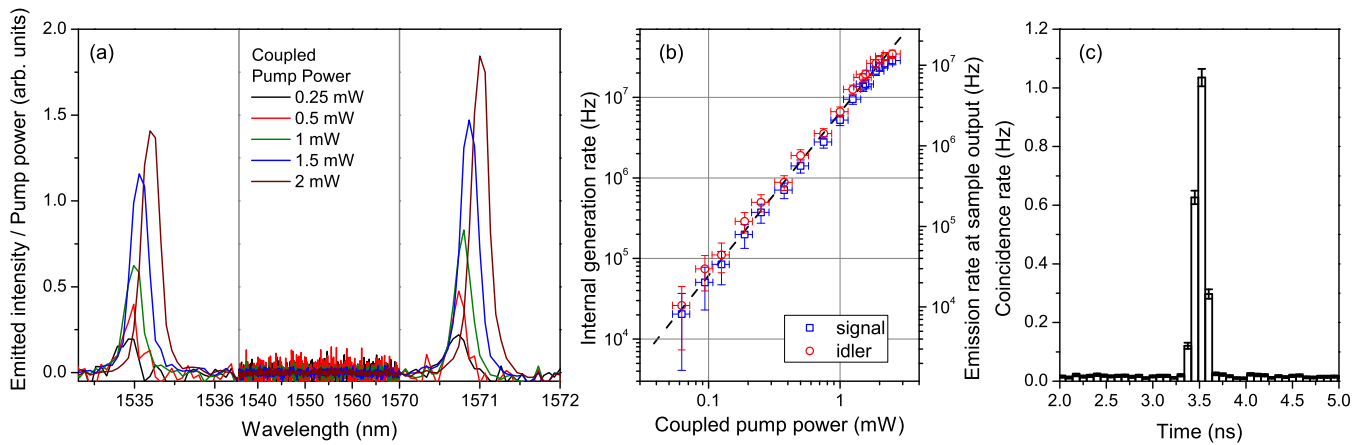}
\caption{\textbf{Spontaneaous FWM and coincidences.} Spectra of the spontaneous FWM experiment for five different
coupled pump powers, signal and idler intensities are divided by the
corresponding pump power to underline the superlinear growth of the
intensity. The slight difference in intensity between the two peaks is due
to slightly different coupling to the input/output bus of the signal and
idler modes. For coupled pump powers above 1 mW, the pump wavelength was
retuned to compensate the red shift of the resonance due to the
thermo-optic effect \protect\cite{Lipson_nat}. The horizontal scale is
expanded around the signal and idler resonances, while the complete absence
of detected photons at the pump resonance confirms the excellent rejection
of the transmitted pump intensity in the set-up. (b) Scaling of the internal generation rates of signal (blue squares) and idler (red circles) photons in
SFWM, varying the coupled pump power. The black dashed line is a guide to
the eye proportional to the square of the pump power. The left axis indicates the photon flux measured at the sample output. (c)
Measured coincidence histogram for a coupled pump power $P_p$ = 1 mW. The time resolution is 75
ps, and it is driven by the response time of the detectors.}
\label{Fig2}
\end{figure*}

Four wave mixing spectra are shown in Fig. 2a: two clear peaks of
generated photons are evident at the signal and idler frequencies. It is important to notice that the pump laser is completely filtered out, so that only spontaneously generated photons are detected. The parametric nature of the emission process is confirmed by the superlinear increase of the generation rate with increasing pumping powers: the quadratic behavior of the generated beams is reported in Fig. 2b, where we plot the estimated generation
rate of photon pairs inside the ring resonator and the output rate \cite{OL}. The output rate was directly measured at the sample output, as is detailed in the Supplementary information. The internal generation rate was estimated in the following way: we have directly measured a total insertion loss of 7 dB for the sample. Due to the sample symmetry, we assume that propagation and coupling losses from the ring resonator to the output fiber to be 3.5 dB. The internal generation rate is then estimated from the flux measured at the output by subtracting the 3.5 dB.

The generation rate can exceed 10$^{7}$ Hz; this is an extremely high rate, and
will be beneficial for all experiments involving coincidence counting. The first step necessary to verify entanglement is to check that signal and idler photons are emitted in pairs: this was assessed via a coincidence experiment, in which the relative times of arrival of idler and signal photons were statistically analyzed \cite{OE}. The coincidence measurement shown in Fig. 2c is obtained employing the set-up described in Fig. 3a (and detailed in Supplementary Fig. 1) by masking the short arm of each interferometer (see Methods). The total losses undergo by signal and idler in the coincidence experiment are 31 dB and 34 dB respectively (Supplementary Note 1).

\begin{figure}[t!]
\includegraphics[width= \columnwidth]{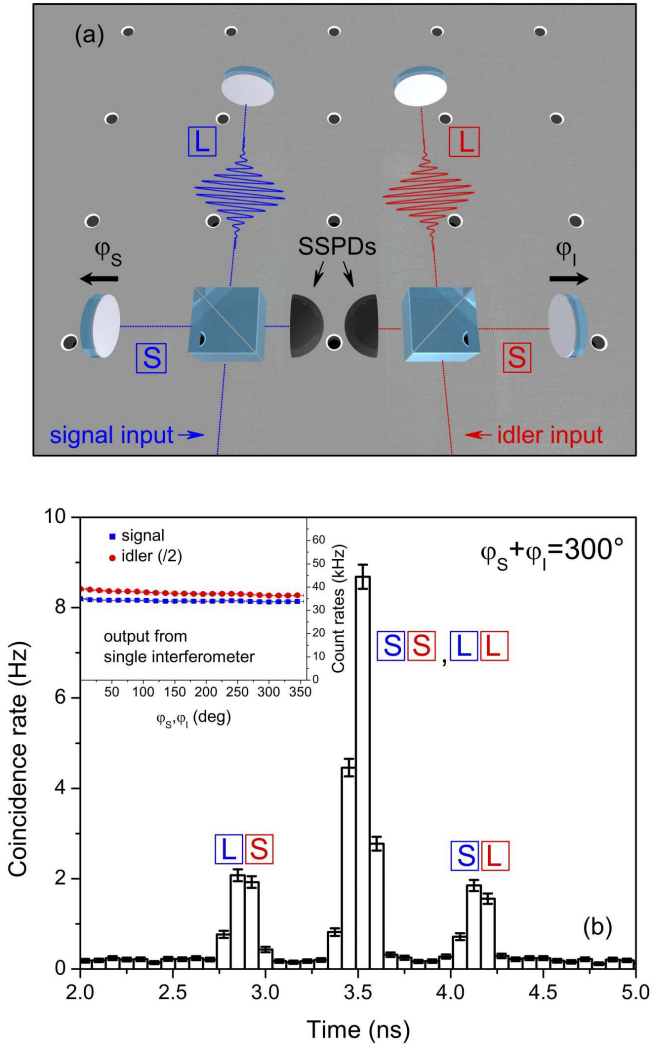}
\caption{\textbf{Correlations at the output of a double interferomenter.} (a) Sketch of the signal and idler Michelson interferometers. The
arm length difference of the two interferometers is the same to well within
the coherence length of the generated photons (Supplementary Note 1). The movable mirrors on the
short arms are connected to a piezo actuator and are used to control the
relative phase between the short and long arms. At the outputs of the
interferometers are two Superconducting Single Photon Detectors (SSPDs). (b)
An instance of coincidence histogram measured at the output of the
interferometers, taken for a couple pump power of 1.5 mW. The integration
time is 120 s. The error bars indicate the error on the counts.
The inset shows the absolute intensity at the output of each interferometer
while varying the respective phase: the complete absence of interference
confirms that the arm length difference is much larger that the coherence time
on the generated photons.}
\label{Fig3}
\end{figure}

An instance of a histogram of the arrival times is shown in Fig. 2c, where a distinct coincidence peak is visible over a small background of accidental counts; this is a clear signature of the concurrent emission of the signal-idler pairs. The 3.5 ns offset is determined by the different path lengths of the signal and idler photons. The coincidence measuments, for experimental consistency, were taken using the same set-up used to measure the entanglement as described below, by masking one arm in each interferometer. The losses from the set-up were directly measured for each component, and amount to 31 dB for signal photons and 34 dB for idler photons, giving a total loss of 64 dB on the coincidence rate. Almost all of these losses, as detailed in the Supplementary Information, are outside the source and are mainly given by the low quantum efficiency of the detectors used in these experiments and the interferometers.

Accidental counts are primarily due to emitted pairs of which only
one photon is detected.  The accidentals in the coincidence curve mainly come from detection of signal and idler photons belonging to different pairs. As all emission times are equivalent, the signal-to-noise ratio (SNR) is also an indication of how likely it is for multiple pairs to be generated at the same time \cite{Gisin_fs}. In our case the SNR is about 70 in Fig. 2c, and higher than 100 in some of the measurements (see Methods and Supplementary Note 3).

\section{Entanglement test on the emitted photon pairs}

The photon pairs are emitted simultaneously but, because of the continuous wave pumping, the emission time is indeterminate to within the coherence time of the pump laser; this is several $\mu$s in our experiments. This systematic lack of information can lead to the pairs being time-energy entangled, as first pointed out by J. D. Franson \cite{FransonPRL}. In order to experimentally measure the entanglement we have used a double interferometer  \cite{FransonPRL,FransonPRA}, as shown in Fig. 3a. Photons at idler frequencies enter one interferometer, while photons at signal frequencies enter the other (see Methods Supplementary Note 1 and Supplementary Note 4). The unbalance $\Delta T$ between the two arms of the interferometers must be much greater than the coherence time $\tau $ of the signal and idler photons to avoid first order interference. In our case $%
\Delta T\sim 0.67$ ns while $\tau \sim 10$ ps ($\tau$ was extracted from the linewidth of the modes). 

\begin{figure}[b!]
\includegraphics[width= \columnwidth]{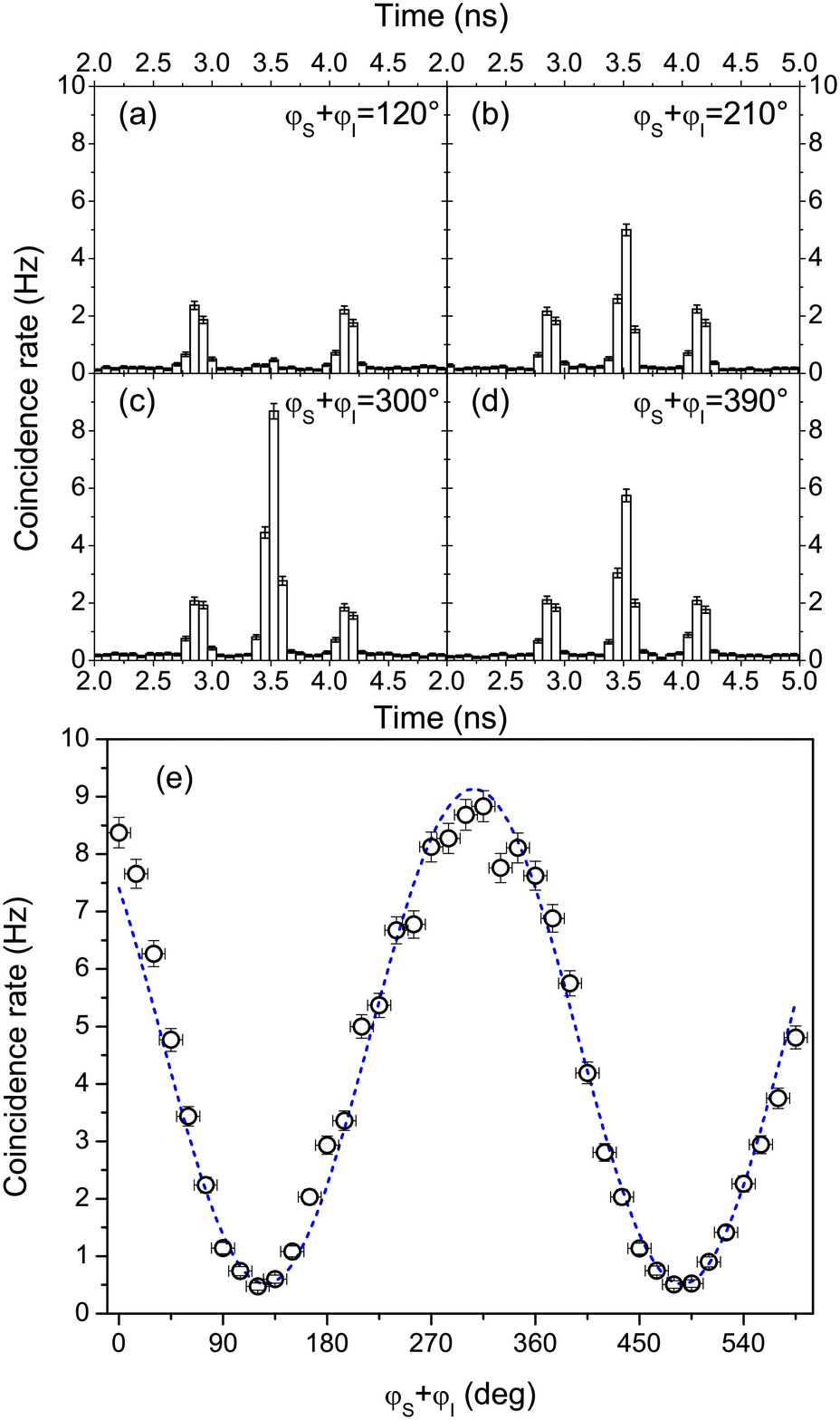}
\caption{ \textbf{Entanglement between signal and idler photons.} (a)-(d) Histograms of the coincidence rate for four different
phase settings. (e) Two photon interference of the double interferometer
configuration: the coincidence count rate of the central peak is plotted
as a function of the phase $\protect\varphi_s + \protect\varphi_i$. The
integration time is of 120 s for each point and the pump power is 1.5 mW.
The dotted black curve is a best fit of the experimental data.}
\label{Fig4}
\end{figure}

The absence of interference in each single interferometer was verified by varying the path differences independently in each of the interferometers and confirming that there was no change in the counting rate detected by the SSPDs, as shown in the inset of Fig. 3b. 
Then with the interferometer arms fixed \cite{Grassani2014} we measured the arrival time of signal photons with respect to idler photons; the generated histograms reveal three relative arrival times. The earlier peak is due to the signal photon having taken the long path in the interferometer and the idler photon having taken the short path; the reverse holds for the latest peak. The middle peak is due to two indistinguishable paths: both photons taking the long path or both taking the short path. The inability to distinguish from which of these two cases the coincidence event arises, due to the long coherence time of the pump, causes second order interference \cite{FransonPRL}. The coincidence rate for the central peak is
expected to be 
\begin{equation}
C\left( \varphi \right) =2C_{0}\left( 1+\cos \left( \varphi +\vartheta
\right) \right),   \label{Franson}
\end{equation}%
where $C_{0}$ is the detected coincidence rate measured by covering one arm in each interferometer. Since signal and idler photons propagate in the same direction once they exit the sample, the phase term $ \varphi $ in the above expression is given by the sum of the phases acquired by the photons passing through the long arms with respect to the short ones: $\varphi =\varphi ^{s}+\varphi ^{i}$; and $\vartheta $ is a constant phase term dependent on the unknown actual lengths of the interferometer arms. 

The effect of varying $\varphi$ is shown in Fig. 4 (the full experimental dataset is shown in Supplementary Fig. 2 and Supplementary Fig. 3).
While the side peaks, corresponding to distinguishable events, have heights that are  independent of $\varphi$, the number of the coincidence counts of the central peak oscillate between minima, close
to zero events, and a maximum, close to four times the height of the side peaks, as shown in Fig. 4a-d. The height of the central peak as a function of $\varphi $ is summarized in Fig. 4e. The trend is well fitted by a sinusoid curve of the type of eq (\ref{Franson}). 
For the data of Fig. 4e, the best fit yields a visibility $%
V_{Meas}=89.3 \%\pm 2.6\%$ $(\mbox{ greather than } 1/\sqrt{2})$, proving a violation of Bell's inequality by
7.1 standard deviations, and so we can conclude that we are generating time-energy entangled photon-pairs \cite{KwiatPRAR2472}.

\begin{table*}
\caption{\textbf{Violation of Bell inequalities.} Summary of the measured parameters for five values
of the coupled pump power $P$. $R$ is the pair emission rate, SNR the signal
to noise ratio, $V_{Meas}$ is the visibility of the two photon interference extracted from the experimental raw data and $\frac{V_{Meas}-1/\sqrt{2}}{\sigma_{V_{Meas}}}$ is the number of standard deviation by which the Bell inequality is violated. Finally, the visibility $V$ is $V_{Meas}$ corrected for the limited visibility $w=0.95$ of the interferometers: $V=\frac{V_{Meas}}{w}$. }
\label{table}
\begin{center}
\begin{tabular}{|l||c|c|c|c|c|c|c|}
\hline
P (mW) & R (MHz) & SNR & $V_{Meas}$ (\%) & $\frac{V_{Meas}-1/\sqrt{2}}{\sigma_{V_{Meas}}}$ & V (\%)\\ \hline\hline
0.25$\pm$0.025 & 0.4 $\pm$ 0.11 & 131.6$\pm$16.5 & 94.8$\pm$3.8 & 6.4 & 99.8$\pm$4 \\ \hline
0.5$\pm$0.05 & 1.7 $\pm$ 0.3 & 120.4$\pm$7.9 & 88.2$\pm$4.8 & 3.6 & 92.8$\pm$5.1 \\ \hline
1.0$\pm$0.1 & 5.8 $\pm$ 0.8 & 64.4$\pm$3.3 & 91.8$\pm$1.9 & 11.2 & 96.6$\pm$2.0 \\ \hline
1.5$\pm$0.15 & 14 $\pm$ 1.9 & 45.1$\pm$2.2 & 89.3$\pm$2.6 & 7.1 & 94.0$\pm$2.7 \\ \hline
2.0$\pm$0.2 & 27 $\pm$ 3.1 & 22.9$\pm$1.0 & 83.8$\pm$3.2 & 4.1 & 88.2$\pm$3.4 \\ \hline
\end{tabular}%
\end{center}
\end{table*}

\section{Discussion}

The experiment was performed for various pumping powers $P$ (Supplementary Fig. 4), and the results are summarized in Table 1 (Supplementary Note 2 and Supplementary Note 3). The Bell's inequality is violated in all cases and by more that 11 standard deviations in the best case. The visibility is limited by the background due to emission of multiple couples and possibly other parasitic luminescent processes (for instance four-wave-mixing and Raman scattering in the access
waveguide and in the optical fibers in the setup): the SNR, as expected, decreases with increasing the pumping power, but it is always sufficiently high to grant entanglement.  It is worth noticing that the values of the visibility $V_{Meas}$ reported in the table are obtained by a single fit operation on the raw data without performing any data correction, e. g. without subtracting the dark counts of the detectors). Finally, the maximum measurable visibility is limited by the first order visibility of the interferometers, in our setup $w=0.95$, which gives the expected visibility $V=V_{Meas}/ w$ (see the last column of Tab. I)

In conclusion, we have experimentally demonstrated a microstructured,
CMOS-compatible source of entangled photons, operating at room temperature with unprecedented capabilities. While ring resonators have long been studied theoretically as a source of quantum correlated states, and pairs of photons emitted from spontaneous four-wave mixing in silicon ring resonators have been detected, 
with this work the oft-quoted promise that these devices
could serve as sources of entangled photons has finally been fulfilled. We confirmed the violation of Bell's
inequality by more than seven standard deviations, and we demonstrated the generation of time-energy entangled photon pairs particularly relevant for telecommunication applications. The source has
incomparable operating characteristics. The spectral brightness per coupled pump power is extremely high, at about 6$\times$10$^{7}$ nm$^{-1}$mW$^{-2}$s$^{-1}$. This is more than four order of magnitudes larger than that reported for entangled photon pairs emitted by long silicon waveguides \cite{Take_Pol,Take,Clemmen}. Even when compared to room temperature sources of entangled photons based on $\chi^{2}$ nonlinearities, which are typically not CMOS-compatible, the emission rate reported here is remarkable. It is two orders of magnitude larger than that obtained from GaAs based waveguides \cite{Ducci} for 1 mW of pumping power, and, for the given bandwidth, it is only a factor of 5-6 less than what can be obtained with \textit{centimeter} long waveguides in periodically poled crystals \cite{Fujii}; our source
has a footprint of less than a hundred square microns. 
This small footprint has great advantages for scalability: all the existing know-how of integrated photonics can be directly used with our source, and its micrometric size makes it ideal for integration with other devices on the same chip, e.g. integrated filters for the pump and routing of signal and idler, both functions well established in integrated photonics. 
Ring resonators are also a well established industrial standard, already used in modulators, for example. Here we have demonstrated a new, compelling functionality of ring resonators: they can be used as sources of entangled states of light. Their production readiness has an immediate large impact for application, much more than other structures characterised by  larger nonlinearities but also a less mature integration  \cite{PhM}.
The signal and idler beams have a bandwidth of $\sim$ 13 GHz, which would allow their use in DWDM network systems without the need of any spectral filtering; the pump powers used here, on the order of dBm, are characteristic of that used in fiber networks; and the pump, signal, and idler frequencies lie in the telecommunications band. We can confidently expect that silicon mircoring resonators will become the dominant paradigm of correlated photon sources for quantum photonics, both for applications involving the transmission of quantum correlations over long distances, such as quantum cryptography, and for applications involving quantum information processing ``on-a-chip''.   

\section{Funding}
This work was supported by MIUR funding through the FIRB "Futuro in Ricerca"
project RBFR08XMVY, from the foundation Alma Mater Ticinensis and by
Fondazione Cariplo through project 2010-0523  Nanophotonics for
thin-film photovoltaics. JES acknowledges support from the Natural Sciences and Engineering Research Council of Canada. MJS and MS acknowledge support from the EPSRC, UK.

\section{Aknowledgements}   
We acknowledge the technical staff of the
James Watt Nanofabrication Centre at Glasgow University.

See Supplementary Information for supporting content.


\newpage
\newpage

\section{Supplementary information for the manuscript ``An integrated silicon source of time-energy entangled photons''}

\subsection{Sample fabrication}

The devices were fabricated on a silicon-on-insulator (SOI) wafer from
SOITEC with a 220 nm-thick silicon core layer. The 5 mm-long and 500 nm-wide
single-mode silicon waveguides for TE-polarized light were fabricated
comprising spot-size converters (SSCs) for efficient input and output
coupling. SSCs consist of 300 $\mu$m-long silicon inverse tapers with a 20 nm
tip width and 1.5 $\times$ 2.0 $\mu$m$^2$ polymer (SU8) waveguides. Ring resonators
with radii of 5, 10, 20 and 30 $\mu$m were fabricated side-coupled to a
single bus waveguide at different coupling distances for each radius. The
pattern of the silicon waveguide structures was defined in negative tone
resist via electronic beam lithography, and then transferred into the
silicon layer by using inductively coupled plasma reactive ion etching.
Polymer waveguide-couplers were written in SU8 material using a second
e-beam lithography step. The silicon chip was finally coated with a
protective PMMA layer. We chose a ring resonator with radius R=10 $\mu$m and
a coupling distance of 150 nm for the experiments reported in this paper
based on its efficiency in the generation of photon pairs.

\section{Detailed experimental set-up}

\begin{figure*} [t!]
\centering
\includegraphics[width=\textwidth]{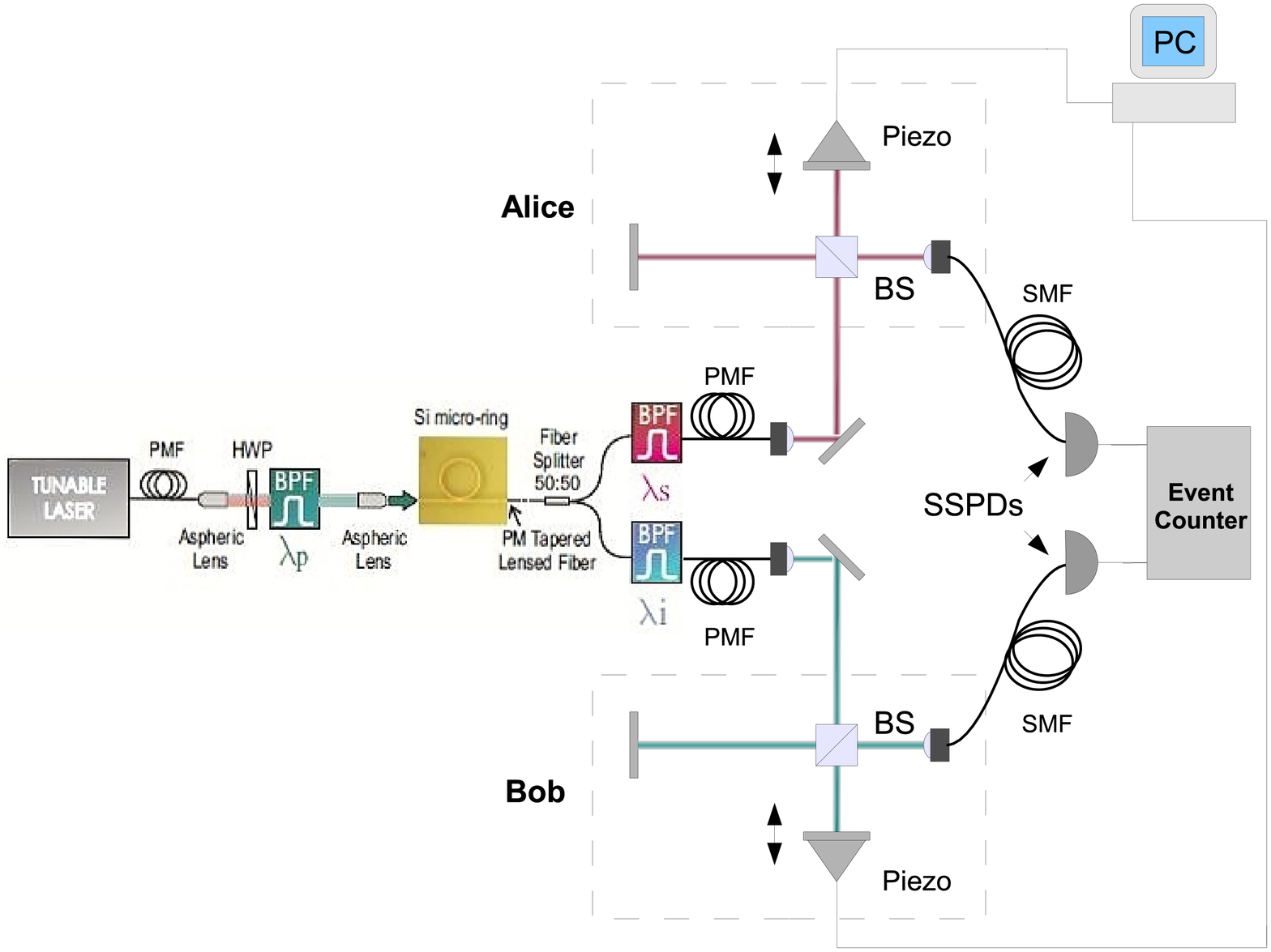} 
\caption{\small{{Complete scheme of the experimental set-up. The acronyms stand for: PMF, polarization maintaining fiber; SMF, single mode fiber; BPF, band-pass filter; FP, fiber polarizer; Piezo, piezoelectric actuator.}}}
\label{Suppl_6e1}
\end{figure*}

Figure \ref{Suppl_6e1}, gives a detailed scheme of the actual setup we employed. Light from a tunable cw pump diode laser (Santec, TSL-510), whose wavelength can range through 1500-1630 nm, comes through a polarization maintaining (PM) single-mode optical fiber.
The pump laser light is collimated to free-space by means of an aspheric lens and then it passes through a half wave plate (HWP) and some pass band interferometric filters (PBFs) to achieve more than 120 dB of side-band suppression at the frequency of the spontaneously generated photons. 
The collimated and filtered pump laser is then injected into the silicon wire waveguide (SWW) by means of a second aspheric lens, which focuses the beam to a diffraction limited spot (diameter $\sim$ 2 $\mu$m) onto the waveguide's edge. The HWP is turned to align the pump polarization to the TE-like single mode SWW. The sample has polymeric field mode converters to couple the silicon wire waveguides to the optical fibers.

The output from the sample is collected by a PM lens tapered fiber mounted on a high precision fiber rotator. In-coupling and out-coupling losses are estimated to be the same, and they sum up to a measured value of about 7 dB. Then two band pass filters (BPFs), tuned at the signal and idler frequencies, are employed at the output of a 50:50 fiber beam splitter in order to separate the signal from the idler. 

\subsection{Characterization and Four Wave Mixing measurements}

Linear transmission spectra with a resolution of 5 pm were taken using a
tunable continuous wave laser modulated through an optical chopper for
injection and an InGaAs photodiode coupled to a lock-in amplifier for
detection. The experimental setup used to test spontaneous four wave mixing
uses a Santec TSL-510 tunable cw laser. Spectral filtering of the laser beam
is performed before injection to filter out the fluorescence background
emitted from the laser because of amplified spontaneous emission. The pump
beam is collimated to free-space using an aspheric lens, its polarization is
aligned to the TE-like single-mode of the silicon waveguide, and finally the
pump light is coupled to the sample using an aspheric lens. The output from
the sample is collected with a PM lensed-fiber mounted on a high-precision
fiber rotator. The total insertion losses of the silicon chip are estimated
around 7 dB. The output containing the transmitted pump beam and the photon
pairs is sent, through a 50:50 fiber beam-splitter, to two band pass filters
(BPF) tuned to the signal and idler frequencies in order to reject the
remaining pump. Using a 50:50 fiber beam-combiner, the output of both band
pass filters is finally conveyed to a monochromator coupled to a
nitrogen-cooled CCD.

\subsection{Measurement of the generation and output rates}

The output rate was measured in an ad-hoc set-up: we calibrated the response of the LN-cooled CCD camera using a power-meter with sub-pW sensitivity, so that the CCD could be then used as a power-meter itself. Between the sample and the CCD we have placed only a filtering stage (to select only signal or idler photons and completely rejecting all other resonances including the pump). The insertion loss of the filtering stage was directly measured to be 3.5 dB. 
To extract the internal generation efficiency, we have back-traced this flux to the ring by adding an additional 3.5 dB of coupling losses (half of the 7 dB insertion loss of the whole sample).

\subsection{Coincidence measurements}

After the signal and idler filters,  collimators couple the photons from the optical fibers to free space. Signal and idler photons then enter the two separate, equally unbalanced, Michelson interferometers. The path length difference is about $\Delta L \simeq 18$ cm, well above the coherence length of the generated photons ($l_c \simeq 2$ mm). The mirrors on the short path of the interferometers are positioned using an open loop piezo actuator (Attocube series 101) with resolution better than 1 nm.  The piezo actuator is connected to a computer-based feedback system that allows the stabilization of the interferometers and the control of their relative phase (see the next section). The collimated photons exiting from the interferometers are then coupled to two single mode fibers by which signal and idler photons are sent to two superconducting single photon detectors from Scontel. 
%

Losses were directly measured for all components of the set-up. In-coupling and out-coupling losses of the sample are estimated to be the same, and they sum up to a measured value of about 7 dB. So the outcoupling losses from the sample are $L_{out}=3.5$ dB, then the beamsplitter introduces another $L_{bs}=4$ dB, while losses from the filters have been measured to be $L_{filter}=3.5$ dB. 
Since we are only collecting from one of the two outputs of each interferometer (with the second output coinciding with the input), half of the signal and idler photons are lost. Moreover, for every interferometer, nonunitary mirror reflectivity, beam splitter imperfections, and coupling losses from fiber to free-space, and viceversa, result in a total loss $L_{Interf}\simeq$10 dB for both the short and the long paths. Detector 1 has a detection efficiency of 10\%, so $L_{D1}=10$ dB, while detector 2 has a detection efficiency of 5\%, so $L_{D2}=13$ dB. 
In total we have a loss $L_s=L_{out}\cdot L_{bs}\cdot L_{filter}\cdot L_{interf}\cdot LD1 = 31$ dB for signal photons, and a loss of $L_i=L_{out}\cdot L_{bs}\cdot L_{filter}\cdot L_{interf}\cdot LD2 = 34$ dB for idler photons. The detected coincidence rate is then given by $R_{coinc}=R \cdot L_s \cdot L_i$ where $R$ is the internal generation rate, so on $R_{coinc}$ there is a loss of about 65 dB.

A very important condition to fulfill in a Franson type experiment is the following relation between the relevant time quantities: $\tau_c \ll \Delta T \ll \tau_p$, where $\tau_c$ is the coherence time of signal and idler photos, $\Delta T = \Delta L/c$ is the time difference between the two paths of each interferometer, and $\tau_p$ is the pump laser coherence time. 
The second fundamental condition is that both interferometers must have the same path difference to within the coherence length of signal and idler photons. In order to align the path difference of the two interferometers we used the following procedure. The first interferometer (at Alice's side) was aligned using a highly coherent laser beam. The second interferometer (at Bob's side) is then roughly aligned (to within about one centimeter) by means of a measuring tape. Then, a pulsed laser (spectral width $\sim$ 8 nm, and pulse coherence time $\sim$ 0.5 ps) is injected in Alice's interferometer whose output is connected in cascade to the input of Bob's device. Finally, the output of Bob's interferometer is sent to a slow detector. Then, the piezo actuator controlling Alice's mirror is scanned with an excursion larger than the laser wavelength. Since the coherence length of the laser pump pulses ($\delta l_p$ $\simeq$ 150 $\mu$m) is much shorter than $\Delta L$, interference arises when $\Delta L_A \simeq \Delta L_B$. Note the interference involves only two of the four paths a pump pulse can take, and thus its visibility is limited to 50 \%. By maximizing the interference's amplitude oscillation, we were able to set the same value of $\Delta L$ for both interferometers to within $\sim150 \mu$m, well below the coherence length on the generated photons (estimated to be about 3 mm).

\subsection{Stabilization of the interferometers}

\begin{figure}[h!]
\centering
\includegraphics[width=\columnwidth]{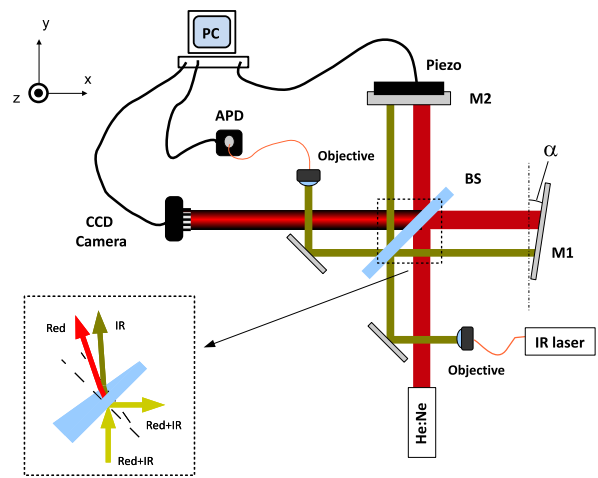} 
\caption{\small{{\upshape Scheme of the stabilized interferometer comprising a control red beam and a target (measured) IR beam. The inset shows the transmission of the red and IR beams through the wedge beam-splitter, the effect is intentionally exaggerate to better show the effect.}}}
\label{set-up interferometro1}
\end{figure}

\begin{figure}[h!]
\centering
\includegraphics[width=\columnwidth]{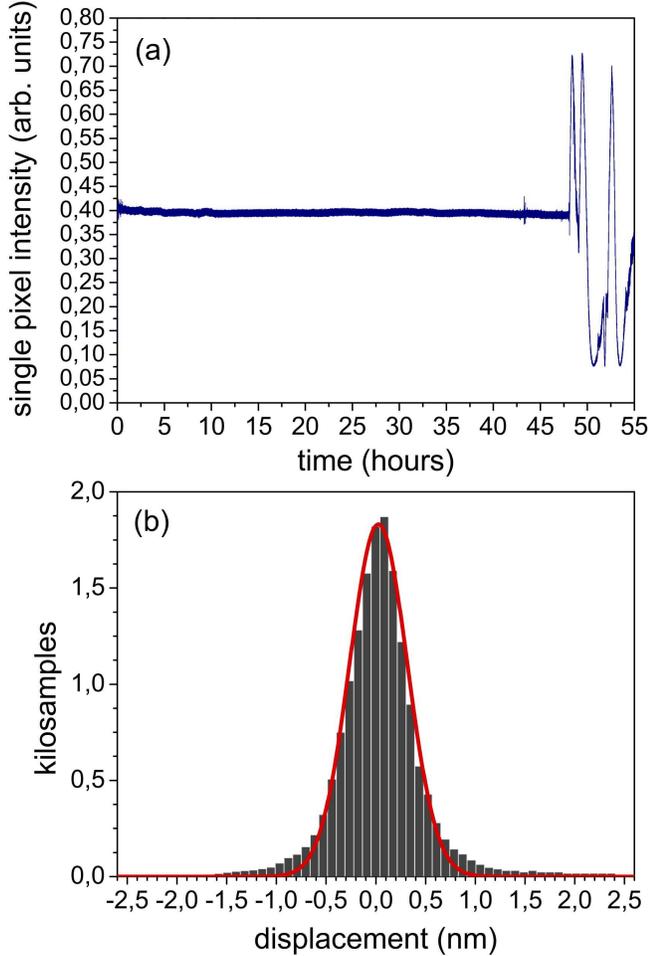} 
\caption{\small{{\upshape (a) Interferometer output with and without the feedback as a function of the acquisition time. (b) Statistical distribution of the phase errors $\delta$ during the stabilization.}}}
\label{Suppl_Interf}
\end{figure}

In order to verify time-energy entanglement between photon pairs produced by our silicon microring resonators, it is necessary to relate the coincidence rate at the output of two separate interferometers, while varying their phases $\phi_A$ and  $\phi_B$.  The period length of the interference is about half the pump wavelength ($\sim 775$ nm). Because of the sub micrometric period of the interference, together with the need of long integration time in coincidence measurements, environmental noise, due for example to thermal fluctuation or mechanical stresses on the optical table, can seriously affect the results of the measurement. This configuration therefore requires very accurate control and stabilization of the path length difference between the arms of each interferometer. To achieve this aim we have developed a new stabilization method by which it is possible to actively stabilize an interferometer at an arbitrary phase with sub-nanometric resolution. The scheme of our set-up is reported in Figure \ref{set-up interferometro1}: a He:Ne laser ($\lambda_P=$632.8 nm) (depicted in red in Figure \ref{set-up interferometro1}) is used as a reference beam, while signal and idler photons follow the controlled IR beam path (in green). One of the two mirrors (M1) is slightly tilted and forms an angle $\alpha$ with the plane P, which is parallel to the wavefront of the incoming light ($y$-direction), so that the two parts of the reference laser beam traveling in the two arms exit the interferometer at a slightly different angle and the interfering beams are no more collinear. This angle results therefore in vertical spatial interference fringes along the $y$-direction.

In order to obtain a collinear geometry on the controlled IR beam while preserving the spatial interference fringes along the $y$-direction on the red reference beam, we injected the reference and controlled beams into the interferometer following two spatially separated parallel directions, and we employed a UV fused silica wedged (wedge angle: 30 arcmin) beam splitter, with an antireflection coating ranging from 600 nm to 1700 nm. The optical dispersion of silica ($dn/d\lambda<0$), together with the wedge angle, cause the transmitted beam of this beam splitter to exit with an angle dependent on its wavelength. Thus, if we adjust the mirror M1 on the transmitted arm (see Figure \ref{set-up interferometro1}) such that it is perpendicular to the IR beam, this mirror reflects the red beam at a certain angle with respect to the incoming wave. Remembering that the mirror on the reflected arm is perpendicular to both beams, this leads to a spatial interference fringe for the reference beam, while the controlled beam conserves a homogeneous phase over its transverse area. In practice, the wedge angle of our beam splitter is sufficient to create about three spatial fringes on the red output beam. 

The fringes on the reference beam are measured using a line CCD camera. The phase of the fringe pattern is directly correlated to the phase of the IR beam, so in essence this methods converts a temporal modulation to a spatial modulation. The camera signal is used as an input for PID control algorithm driving the piezo actuators in the interferometers. We have checked that the interferometers can be set and stabilized at an arbitrary phase. An example of stabilization is shown in Figure \ref{Suppl_Interf}, where the phase has been kept stable for 48 hours and then the PID control algorithm was switched off. Analysis of the data confirms that both stability and positioning accuracy are about 0.8 nm, limited by the piezo positioning accuracy. This corresponds to a precision in phase of about one third of a degree, for the signal and idler wavelengths.

A full description of the stabilization method and further details can be found in Grassani \emph{et al.} Optics Letters 39, 2530 (2014)..

%
%
%

\subsection{SNR and visibility from the data}

Levenberg-Marquardt nonlinear regression of the function
\begin{equation}
\label{sinefunction}
F\left(\varphi,\left(A,\vartheta,y_0\right)\right)=A*\cos\left(\varphi+\vartheta\right)+y_0
\end{equation} 
to the experimental data was performed using Mircocal Originlab plotting and analysis software; here $\varphi$ is the sum of the phases of the two interferometers, while $A$, $\vartheta$ and $y_0$ are the fit parameters.
The measured visibility is extracted from the best fit of the experimental data:
\begin{equation}
V_{Meas} = \frac{max-min}{max+min} = \frac{A}{y_0},
\label{Visibility}
\end{equation}
where $max$ and $min$ are respectively the maxima and the minima of the curve.
Since the $min$ value defined in (\ref{Visibility}) depends on the accidental counts of the measure, and the $max$ to the coincidences, we can express $V_{Meas}$ as a function of the signal to noise ratio SNR as follows:
\begin{equation}
V_{Meas} = w \frac{\mbox{SNR}-1}{\mbox{SNR}+1},
\label{Vis_Car}
\end{equation}
where $w=0.95$ is the mean first order interference visibility of the interferometers, measured using a high coherence laser beam. 
The SNR can itself be extracted from the best fit parameters, $\mbox{SNR}=y_0-A$. SNR can however be directly extracted from the measured coincidence curves: as the side peaks of the coincidence curves do not depend on $\varphi$, we can define
\begin{equation}
\mbox{SNR} = 2\times\frac{C_{LS}+C_{SL}}{C_{A}},
\label{SNR}
\end{equation}
where $C_{LS}$ is the count rate for events in which the signal photon has taken the long path and the idler photon has taken the short path in the interferometers (vice versa per $C_{SL}$), while $C_A$ is the accidentals count rate. We have chosen to extract the SNR directly from the data as outlined equation \ref{SNR}: the obtained values are consistent with the SNR measured in direct coincidence experiments (i.e. without interferometers).

\subsection{Error analysis}
The errors are taken into account using standard error propagation theory. The statistical distribution of coincidence counts, including $C_{LS}$, $C_{SL}$ and $C_{A}$, is assumed to be Poissonian. The interferometers stabilization procedure lets us set the value of $\varphi$ to better than 0.4$^{\circ}$, nonetheless we estimate the error on $\varphi$ to be about 10$^{\circ}$ due to phase inhomogeneity on the beam profiles.
The uncertainty on the derived quantities $A$ and $y_0$ is obtained from the fitting routine.
The error on the visibility is thus:
\begin{equation}
\label{sigmaV}
\sigma_{V_{Meas}}=V_{Meas}\sqrt{\left(\frac{\sigma_A}{A}\right)^2+\left(\frac{\sigma_{y_0}}{y_0}\right)^2}.
\end{equation}
The error on the SNR is
\begin{equation}
\label{sigmaSNR}
\sigma_{\mbox{SNR}}=\mbox{SNR}\sqrt{\left(\frac{\sqrt{\sigma_{C_{LS}}^2+\sigma_{C_{SL}}^2}}{C_{LS}+C_{SL}}\right)^2+\left(\frac{\sigma_{C_A}}{C_A}\right)^2}.
\end{equation}

\subsection{The full experimental dataset of figure 4 (e)}

In the following two pages are all the coincidence curves measured for a coupled pump power of 1.5 mW (Fig. 4 in the main text). The curves are reported as measured from the event timer, without any treatment of the data (and thus with no error bars). The time window of 75 ps coincides with the time resolution of the detectors.
\newpage
\begin{figure}[h!]
\centering
\includegraphics[width=\textwidth]{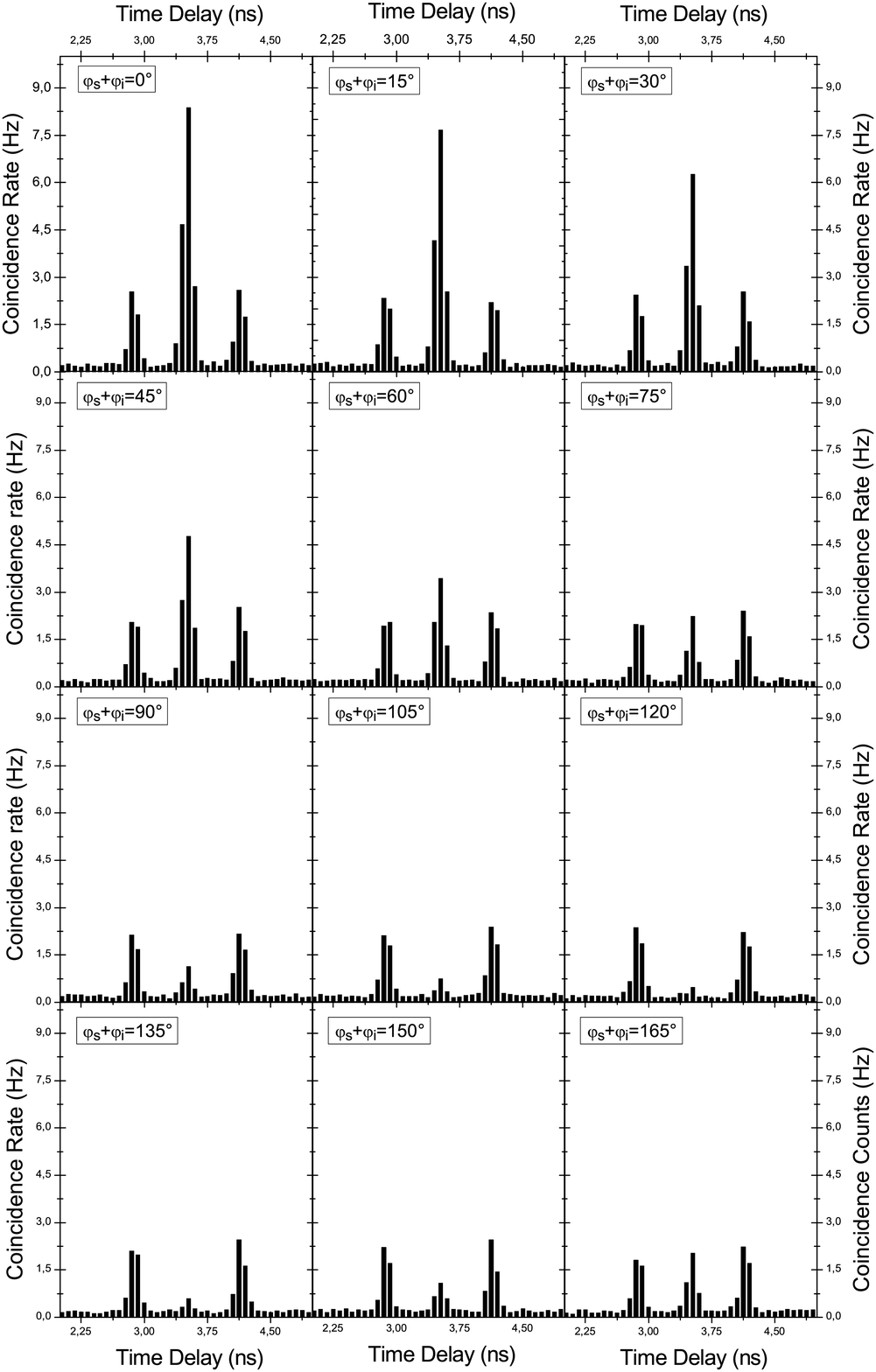} 
\label{Franson_setup}
\end{figure}
\newpage
\begin{figure}[h!]
\centering
\includegraphics[width=\textwidth]{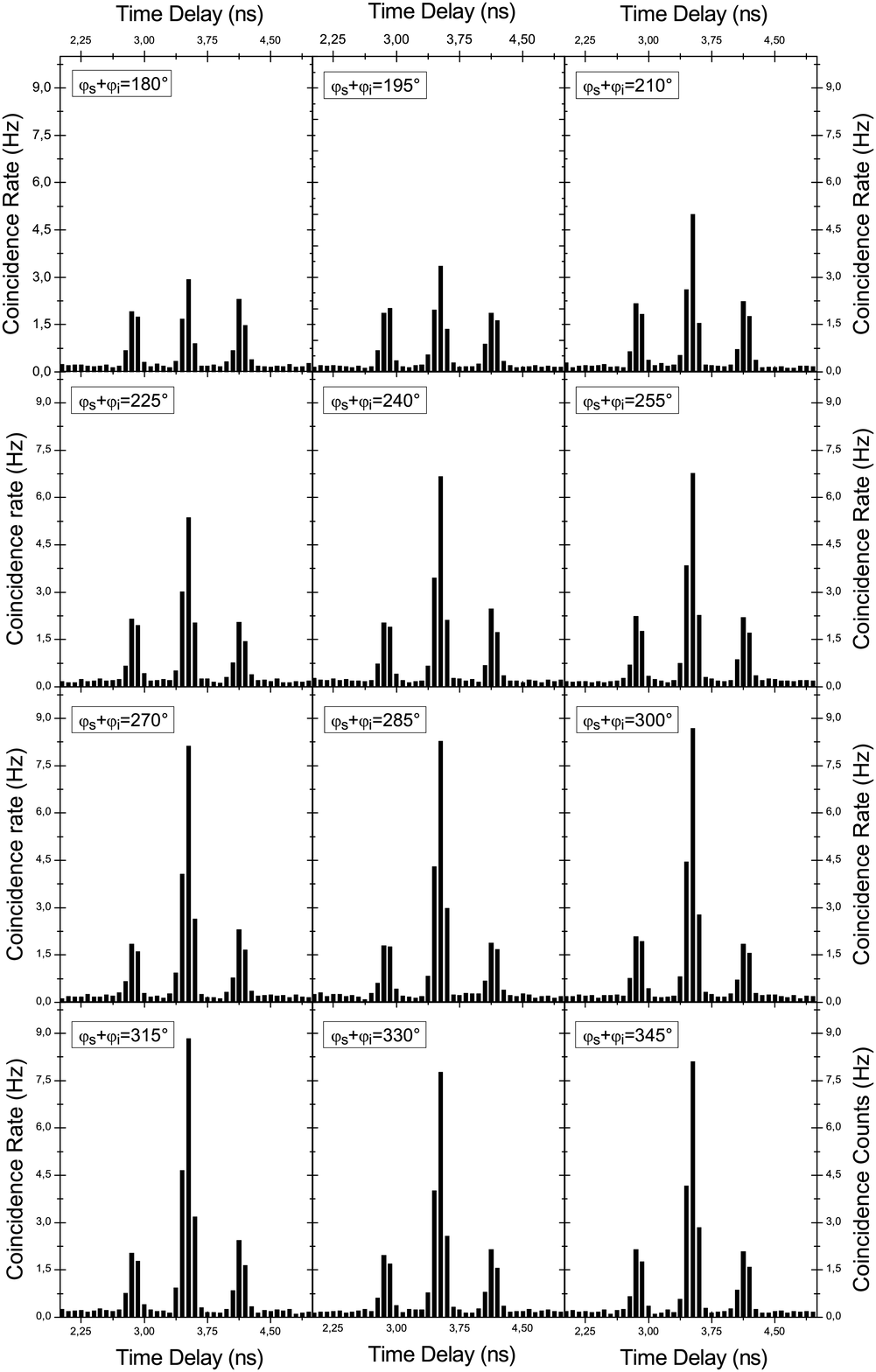} 
\label{Franson_setup}
\end{figure}

\end{document}